\documentclass[twocolumn,showpacs,preprintnumbers,amsmath,amssymb,prl]{revtex4}

\usepackage{graphicx}
\usepackage{dcolumn}
\usepackage{bm}

\begin{document}

\preprint{APS/123-QED}

\title{Chaos assisted instanton tunneling in one dimensional perturbed periodic potential}

\author{V.I. Kuvshinov}
\email{kuvshino@dragon.bas-net.by}
\author{A.V. Kuzmin}
\email{avkuzmin@dragon.bas-net.by}
\author{R.G. Shulyakovsky}
\email{shul@dragon.bas-net.by} \affiliation{Institute of Physics,
National Academy of Sciences of Belarus, \\
Scarina av., 68, 220072 Minsk, BELARUS.}

\date{\today}

\begin{abstract}
For the system with one-dimensional spatially periodic potential
we demonstrate that small periodic in time perturbation results in
appearance of chaotic instanton solutions. We estimate parameter
of local instability, width of stochastic layer and correlator for
perturbed instanton solutions. Application of the instanton
technique enables to calculate the amplitude of the tunneling, the
form of the spectrum and the lower bound for width of the ground
quasi-energy zone.
\end{abstract}

\pacs{03.65.Xp, 05.45.Mt}

\maketitle

\section{Introduction}\label{sec1}
Tunneling as inherently quantum phenomenon attracts much attention
\cite{tunneling}. Its connection with classical chaos in
semiclassical regime has also been discussed
\cite{GeneralChaos,GeneralChaos2}. A number of works were devoted
to semiclassical chaos assisted tunneling between symmetry related
KAM-tori in systems with mixed dynamics (well developed chaotic
region coexists in phase space with regular islands)
\cite{CAT,CAT2,CAT3}. To describe chaos assisted tunneling in
systems with mixed dynamics multi-level model Hamiltonian,
primarily proposed in \cite{GeneralChaos}, is often used
\cite{Example}. Less attention has been payed to semiclassical
tunneling in KAM systems (chaotic region is not
widespread)~\cite{Chemistry}. Another way to describe
semiclassical tunneling is based on solutions of Hamilton
equations in imaginary time and path integral formalism
\cite{tunneling}.  Instanton technique~\cite{BPST} was used in a
very few works \cite{InsCh}.

In this work we consider one-dimensional quantum system with
periodic in space potential affected by small periodic in time
perturbation. We use methods created to describe chaos in
classical Hamiltonian systems to investigate essentially quantum
phenomenon of tunneling. It is achieved in the framework of
instanton technique, where solutions of Euclidian equations of
motion (instantons) play dominating role, by the use of standard
methods from the viewpoint of chaos \cite{Chaos}. For the systems
with periodic in time perturbation energy is no more an exact
integral of motion and the language of quasi-energies is more
adequate~\cite{Floquet}. For some estimations energy as an
adiabatic invariant can also be used \cite{Sagdeev}. We study
properties of chaotic instanton solutions and calculate the form
of the spectrum and the lower bound for the width of the ground
quasi-energy zone.

Hamiltonian of the system under consideration is taken in the form
\begin{equation}\label{1}
\tilde{H} = \frac{1}{2}\tilde{p}^{2} + \omega_{0}^{2}\cos{x} -
\epsilon x \sum_{n=-\infty}^{+\infty} \delta(t - n\tilde{T}),
\end{equation}
$\tilde{T}$ is the real time period of perturbation, $\epsilon$
describes the strength of perturbation. The mass of the particle
equals unit. Considered cosine potential corresponds  nonlinear
oscillator. Phase space of nonlinear oscillator has the topology
of cylinder (points $(q,\tilde{p})$ and $(q + 2\pi, \tilde{p})$
are identified). Thus its (quasi-)energy spectrum is discrete.
Chaos assisted tunneling between two major resonance islands
(inside single potential well) for driven nonlinear oscillator has
been studied numerically in \cite{CAT}. In this work we study the
system with $x$ varying from $-\infty$ to $+ \infty$. This results
in band structure of the (quasi-)energy spectrum~\cite{Bloch}.
Systems with spatially periodic potential were studied in the
instanton physics \cite{Rajar}. Perturbation used in (\ref{1}) was
exploited in the systems exhibiting quantum chaos
\cite{Zaslavsky,Physica}.

There are papers devoted chaos assisted tunneling where some
analytical predictions for billiard systems basing on the path
integral formulation of quantum mechanics are made
\cite{GeneralChaos2}. Distinguishing feature of our work is
analytical predictions for the system with smooth potential
exploring for this purpose instanton technique adopted from
quantum field theory \cite{BPST}.

\section{Analysis of chaotic instanton solutions}

For applying instanton technique we consider solutions of
classical equations of motion in {\it imaginary} (Euclidian) time.
Hamiltonian (\ref{1}) has the same form (translated on $\pi$) in
Euclidian time as in real one.

Euclidian Hamiltonian of the system is $H=H_0+V$, where
\begin{equation}\label{H0}
H_{0} = \frac{1}{2}p^{2} - \omega_{0}^{2}\cos{x},
\end{equation}
and
\begin{equation}\label{perturbation}
  V = \alpha T x \sum_{n=-\infty}^{+\infty}
  \delta(\tau-nT).
\end{equation}
Here $H_{0}$ is nonperturbed Euclidian Hamiltonian of the system
and $V$ is the Euclidian potential of the perturbation. We also
introduced coupling constant $\alpha \ll 1$ instead of $\epsilon
\equiv \alpha T$ in order to simplify formulas.

Nonperturbed instanton solution describes the motion on the
separatrix of the Hamiltonian (\ref{H0}). It is well known that
this separatrix is destroyed by any periodic perturbation and on
its place stochastic layer is appeared \cite{Sagdeev}. Perturbed
instanton solutions correspond to the motion in vicinity of the
separatrix inside the layer. Therefore instead of one instanton
solution connecting neighbor maxima of nonperturbed Euclidian
potential (classical vacuum states in real time potential) we
obtain a manyfold of instanton solutions of Euclidian equations
placed inside the stochastic layer.

We calculate parameter of local instability, width of stochastic
layer and correlator for perturbed instanton solutions. It is
convenient to describe dynamics of the system in action-angle
variables~\cite{Chaos}. Equation of motion for action variable has
the form
\begin{equation}
  \dot{I} = -\frac{\alpha \dot{x}}{\omega} \left( 2 \sum_{m=1}^{+\infty} \cos{(m\nu
  \tau)} + 1 \right).
\end{equation}
Here $\omega (I) \equiv d H_{0}/dI$ is the nonlinear frequency
\cite{Chaos}.  Instead of angle variable we introduce phase of
external force $\psi$ defined by the relation $\dot{\psi} = \nu
\equiv 2\pi / T$ \cite{Sagdeev}. Let $H_{s} \equiv \omega_{0}^{2}$
to denote the energy of nonperturbed system on the separatrix.
Continuous equations of motion for $I$ and $\psi$ can be reduced
to discrete mapping for the phase of external force in the
vicinity of separatrix ($|H-H_{s}|\ll
1$)~\cite{Sagdeev,Zaslavsky,Physica}
\begin{equation}\label{map}
\psi_{n+1} = \psi_{n} + B_{n} + K_{0} \sin{\psi_{n}},
\end{equation}
where $$ K_{0} = \frac{8\pi \alpha\nu}{\omega_{0}}\frac{e^{-\pi
\nu/2 \omega_{0}}}{|H - H_{s}|}, $$ $B_{n}$ are some functions of
$H$ which exact form is not essential for our purposes. We assume
following \cite{Sagdeev} that due to small value of perturbation
the energy practically does not change with time and equals the
energy of the nonperturbed system. The map (\ref{map}) with
arbitrary parameter $K_{0}$ was studied by many authors, for
instance \cite{Zaslavsky}. Particulary, it is known that at
$K_{0}>1$ motion is locally unstable and chaotic, whereas at
$K_{0}\leq 1$ it is stable and regular. Thus $K_{0}$ is the
parameter of local instability. Condition $K_{0}\sim 1$ enables us
to calculate the width of stochastic layer
\begin{equation}\label{layer}
  |H_{s} - H_{b}| = \frac{8 \pi \alpha \nu}{\omega_{0}} e^{-\pi
  \nu/2\omega_{0}},
\end{equation}
here $H_{b}$ is the energy value on the bound of stochastic layer.

To calculate correlator for perturbed instanton solutions we use
standard technique~\cite{Corr}. For the map (\ref{map}) correlator
is
\begin{eqnarray}
  R(\tau, \tau_{0})&&= \frac{1}{2\pi}
  \int_{0}^{2\pi} d \psi_{0} \exp{\{i(\psi(\tau) -
  \psi_{0})\}}\nonumber\\
  &&\sim\exp\left( -\frac{\tau - \tau_{0}}{\tau_{R}}\right),
\end{eqnarray}
here $\psi_{0} \equiv \psi(\tau_{0})$ and the time of correlations
decay is $\tau_{R} = 2\pi/(\omega \ln{K_{0}})$. Exponential
decrease of correlator shows that dynamics of the instanton
solutions inside the stochastic layer ($K_{0} >1$) possesses the
property of mixing (chaos) \cite{Chaos}.

Note that perturbed one-instanton solution due to stochastic layer
connects not only neighbor vacua of real time potential but also
two {\it arbitrary} chosen vacua. Remind that to describe
tunneling between non-neighbor vacua of nonperturbed system one
has to take into account contribution of multi-instanton
configurations~\cite{Rajar}.

\section{The  calculation of the tunneling amplitude and ground zone width}

Let us consider the tunneling between neighbor vacua (from
$x\approx-\pi$ to $x\approx\pi$ for distinctness) in presence of
perturbation (\ref{perturbation}). In Euclidian time this
tunneling process for the nonperturbed system is described by the
solution of Euclidian equations of motion with asymptotes $x=-\pi,
\quad p=0$ at $\tau=-\infty$ and $x=\pi,\quad p=0$ at
$\tau=+\infty$. There is only one solution satisfying these
conditions for nonperturbed system (\ref{H0}) (one-instanton
solution)
\begin{equation}\label{instanton}
 x_{0}^{inst}(\tau - \tau_0)=-\pi+4\arctan{e^{\omega_{0}(\tau-
\tau_0)}}.
\end{equation}
Its Euclidian action is $S^{inst}=8\omega_{0}$. The instanton's
position is denoted by $\tau_0$. Due to Euclidian equations of
motion and anti-symmetry of $x_{0}^{inst}$ when time is inverted
with respect to the point $\tau_0$ perturbation
(\ref{perturbation}) does not change the Euclidian action of the
one-instanton solution (\ref{instanton}) in the first order on the
coupling constant $S^{inst}_{pert} = S^{inst} + O(\alpha^{2})$.
The only manifestation of the perturbation in this approximation
is the appearance of a number of the additional solutions of
Euclidian equations of motion with energies close to the energy of
nonperturbed one-instanton solution and placed inside the
stochastic layer.

Let us consider firstly nonperturbed system at arbitrary energy
$-\omega_0^2+\varepsilon,\quad 0<\varepsilon<2\omega_0^2$. One
half of truncated instanton action can be easily calculated $$
S[x^{inst}(\tau,\varepsilon)]=\int\limits_{-a(\varepsilon)}^{a(\varepsilon)}
\sqrt{2(\omega_0^2\cos x-(-\omega_0^2+\varepsilon))}\, dx= $$
\begin{equation}\label{nonzero}
=4\sqrt{4\omega_0^2-2\varepsilon}E\left(a(\varepsilon),
\frac{1}{1-\frac{\varepsilon}{2\omega_0^2}}\right),
\end{equation}
where $\pm a(\varepsilon)=\pm
\arcsin\sqrt{1-\frac{\varepsilon}{2\omega_0^2}}$ are turning
points, function $E$ is the elliptic integral of the second kind.

Then tunneling amplitude in {\it perturbed} system can be found by
integration over energy of the tunneling amplitude in {\it
nonperturbed} system with the action (\ref{nonzero})
\begin{equation}\label{A2}
A=\int\limits_0^{\Delta
H}d\varepsilon\!\!\!\int\limits_{x(\tau)\approx
-\pi}^{x(\tau)\approx\pi}Dx\exp\left[-S[x^{inst}(\tau,\varepsilon)]\right],
\end{equation}
where $\Delta H=2|H_{s} - H_{b}|$ is the stochastic layer width.
The contribution of the chaotic instanton solutions is taken into
account by  means of integration over $\varepsilon$. Expression
(\ref{A2}) shows that the probability of tunneling (square of the
absolute value of the tunneling amplitude) grows while chaotic
region spreads ($\Delta H$ increases).

The result is obtained in the first order on coupling constant
$\alpha$ and does not take into account possible structure of
stochastic layer is valid if the layer is narrow and is in
agreement with results of numerical \cite{Lin} and real
\cite{CAT3} experiments for similar problems. We have also
correspondence in (\ref{A2}) with the nonperturbed
case~\cite{Rajar}. Namely, if $\alpha =0$ then $\Delta H = 0$ and
the single solution describing the motion on the separatrix
(nonperturbed one-instanton solution) contributes to the tunneling
amplitude.

Formula (\ref{A2}) can be made more transparent if we use the
approximate form of action (\ref{nonzero}) at
$\varepsilon<2\omega_0^2$
\begin{equation}\label{approx_act}
S[x^{inst}(\tau,\varepsilon)]\approx
8\omega_0-\frac{\pi\varepsilon}{\omega_0}.
\end{equation}
Then in Gauss approximation we obtain the following expression for
the tunneling amplitude
\begin{widetext}
\begin{equation}\label{fin}
A=\int\limits_{0}^{\Delta
H}d\varepsilon\int\limits_{-\infty}^{+\infty}dc_0
\sqrt{S[x^{inst}(\tau,\varepsilon)]}\exp\left(
-S[x^{inst}(\tau,\varepsilon)]\right)\approx
e^{-S^{inst}}\sqrt{S^{inst}}\Gamma F=\sqrt{8\omega_0}\Gamma
e^{-8\omega_0}e^{\frac{\pi\Delta H}{\omega_0}}.
\end{equation}
\end{widetext}
where integration over $c_{0}$ gives the contribution of zero
modes, $\Gamma$ is a time of the tunneling. Formula (\ref{fin}) up
to the factor $i$ has the same form in real Minkovski time.

Expression (\ref{fin}) can be interpreted in the following way.
Factor $F=\exp\left(\frac{\pi\Delta H}{\omega_0}\right)>1$ in
(\ref{fin}) differs perturbed and nonperturbed amplitudes and
includes the contribution of the layer. One can think about $F$ as
a number of instanton solutions inside the stochastic layer.

We can find the minimal number of instanton solutions inside the
stochastic layer. One can not distinguish instanton solutions
within energy interval $\Delta E \sim 1/\Delta \tau$ (Heisenberg
uncertainty relation). Here $\Delta \tau\sim \omega_{0}^{-1} $
denotes the time interval of observation. Thus the energy interval
between neighbor instantons is $\Delta \varepsilon \sim
\omega_{0}$. Therefore parameter $F$ can be found as follows
\begin{equation}\label{F1}
F\sim 1+\frac{\Delta H}{\Delta\varepsilon}\approx e^{\frac{\Delta
H}{\omega_0}},
\end{equation}
where unit takes into account the nonperturbed separatrix.

The form of the spectrum of the lower quasi-energy zone is
obtained using the amplitude (\ref{fin}) by means of standard
technique (multi-instanton contributions are taken into
account)~\cite{Rajar}:
\begin{equation}\label{zone}
  E_{\theta}\approx\frac{1}{2}\omega_0- 2e^{-S^{inst}}\sqrt{S^{inst}}F\cos{\theta},
\end{equation}
here continuous variable $\theta$ parametrizes levels of the
ground quasi-energy zone. Zone width
\begin{eqnarray}\label{spectrum}
\Delta E\approx 4e^{-S^{inst}}\sqrt{S^{inst}}F
\end{eqnarray}
differs from the case of nonperturbed case by factor $F$
reflecting the influence of perturbation.

\section{Conclusion}

We applied theory of {\it classical} chaos for investigation of
the chaos assisted {\it tunneling} in terms of path integral
formalism in imaginary time and instanton technique. We found the
parameter of local instability and the width of the stochastic
layer. Exponential decrease of the correlator for any perturbed
instanton solution was also demonstrated, that means it to possess
the property of mixing. Then properties of the stochastic layer
and classical chaotic solutions in Euclidean space (chaotic
instantons) were used for the calculation tunneling amplitude,
ground quasi-energy zone spectrum in the presence of the
perturbation and the zone width.

General tendency for chaos assisted tunneling regime (in average
-- if we abstract from fluctuations) is the increase of tunneling
amplitude (probability) as the strength of perturbation increases
\cite{CAT3,Lin}. It is confirmed here. The reason is the growth of
the width of the chaotic layer and therefore the increase of the
number of paths for particle to travel from one regular region to
another. For small energies in Gauss approximation tunneling
amplitude is increased by the factor $F>1$. The life-time of the
particle in the certain vacuum of the system decreases. It is
connected with the widening of the (quasi-)energy zone
(\ref{spectrum}).

We would like to emphasize that obtained results are not
consequences of the particular choice of the nonperturbed
Hamiltonian (\ref{H0}) or perturbation (\ref{perturbation}).
Qualitatively they are valid for more general class of
one-dimensional nonperturbed Hamiltonians with quadratic
dependence on momentum and spatially periodic potential with
single well in each period, as well as time dependence of
homogeneous perturbation can be realized by any time periodic
function. The reason is the universality of the separatrix
destruction mechanism in these potentials affected by
time-periodic perturbation \cite{Sagdeev}.

Tunneling plays an important role in gauge field theories
(instanton physics~\cite{BPST}). Experimental discovery of QCD
instantons for example is an important problem~\cite{22}. Moreover
it is known that classical gauge field theories are inherently
chaotic~\cite{20}. Therefore the study of chaos assisted instanton
tunneling in gauge field theories based on chaos criterion in
quantum field theory~\cite{23} is also of essential interest.

{\it Acknowledgements.} One of us (A.V.K.) gratefully acknowledges
the grant of the World Federation of Scientists, National
Scholarship Programme -- Belarus.

\end{document}